\documentstyle[12pt,epsfig]{cernart}
\def\la{\mathrel{\mathpalette\fun <}}
\def\ga{\mathrel{\mathpalette\fun >}}
\def\fun#1#2{\lower3.6pt\vbox{\baselineskip0pt\lineskip.9pt
\ialign{$\mathsurround=0pt#1\hfil##\hfil$\crcr#2\crcr\sim\crcr}}}

\newcommand{\bc}{\begin{center}}
\newcommand{\ec}{\end{center}}
\newcommand{\bd}{\begin{displaymath}}
\newcommand{\ed}{\end{displaymath}}
\newcommand{\be}{\begin{equation}}
\newcommand{\ee}{\end{equation}}
\newcommand{\ba}{\begin{array}}
\newcommand{\ea}{\end{array}}
\newcommand{\bt}{\begin{tabular}}
\newcommand{\et}{\end{tabular}}

\begin{document}
\large
~
\vspace{1cm}

\bc
{\Large\bf PION DOUBLE CHARGE EXCHANGE ON NUCLEI\\[0.6cm] AND
INELASTIC RESCATTERINGS}\\[1cm]
{\Large A.B.Kaidalov, A.P.Krutenkova}\\[1cm]
{\Large\it Institute of Theoretical and Experimental Physics \\
Moscow, 117259, Russia}\\[1cm]
\ec
\begin{abstract}
The inclusive pion double charge exchange (DCX) process on nuclei
is considered in the framework of the Gribov--Glauber approach. It is
shown that inelastic rescatterings related to production of two (and
more) pions in the intermediate state give an important contribution
to the process of DCX at kinetic energies above $\sim0.5\;GeV$. This
mechanism dominates at energies $\ga1\;GeV$ and allows to explain the
weak energy dependence of the DCX cross section observed
experimentally [1].

\end{abstract}
\vspace{0.5cm}

\begin{figure}[bhtp]
\begin{center}
\vspace*{1.0cm}
\mbox{\epsfig
{figure=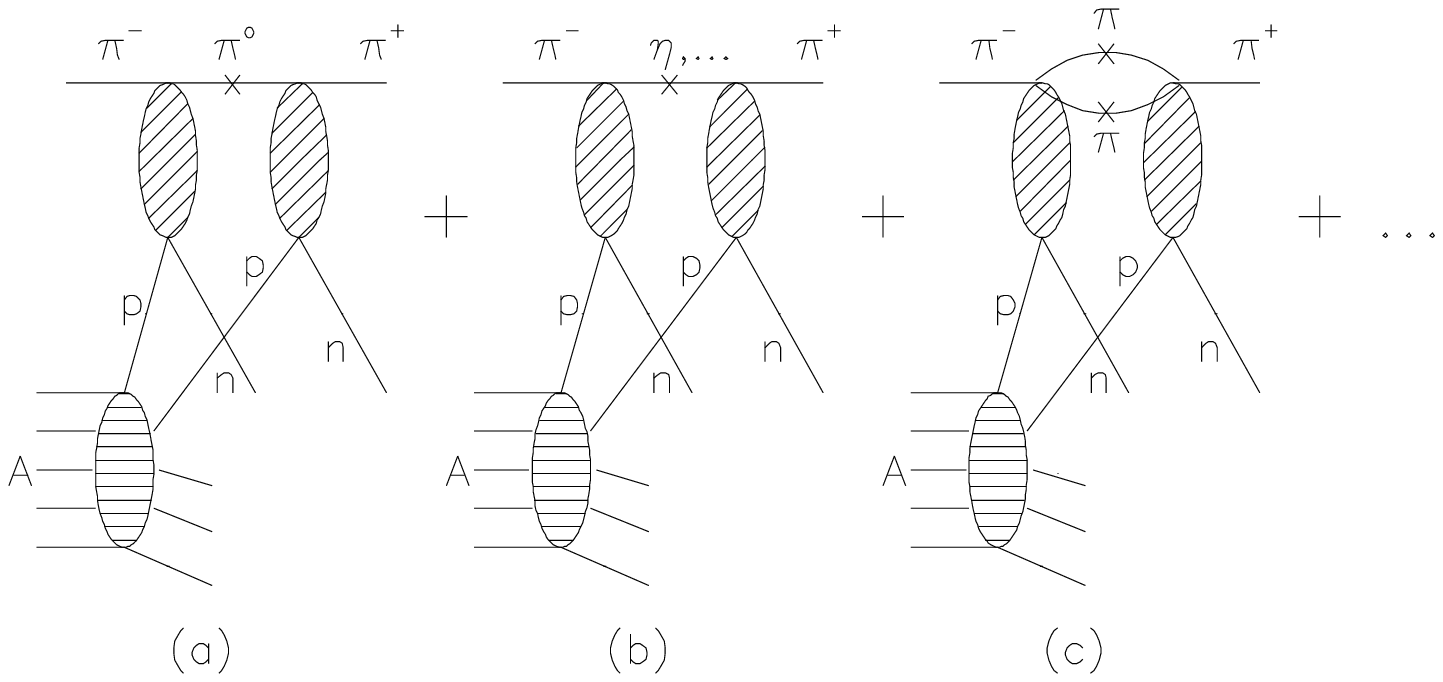,width=11cm,
            bbllx=0,bblly=0,bburx=567,bbury=310}}
\caption [] {Diagrams contributing to pion double charge exchange
on nuclei:

(a) sequential single charge exchanges (SSCX), i.e. standard
mechanism (elastic rescattering),

(b) quasielastic rescatterings,

(c) inelastic rescatterings.}
\label{diagram1}
\end{center}
\end{figure}

The pion double charge exchange process is an interesting probe
[2] of nuclear structure and of a hadron dynamics as it
involves interaction with at least two nucleons of a nucleus. The
standard DCX mechanism corresponds to two sequental single charge
exchanges of pion on nucleons of a nucleus (fig.~\ref{diagram1}(a)). 
This model gives
a reasonable description of experimental data on DCX at kinetic
energies of incident pions $T_0\la0.5\;GeV$ and predicts [3] a
strong decrease of a small angle DCX at higher energies. However
recent measurements of DCX at energies up to
$1.1\;GeV$ performed at ITEP showed [1] that the differential cross
section of this process decreases with energy rather weakly and
exceeds the theoretical prediction [4] at the highest energy by an
order of magnitude.  Thus other mechanisms of DCX are needed to
explain the observed energy dependence.

In this note we will show that inelastic rescatterings of the type
shown in fig.~\ref{diagram1}(b),(c) 
give an important contribution to the cross section
of DCX at energies $T_0>0.5\;GeV$ and allow to understand the
experimentally observed pattern of the energy dependence for this
reaction. In the following we shall use the Gribov formalism for
a description of rescatterings on nuclei [5]. We will also compare
our results with the predictions of a model based on meson exchange
currents.

The inclusive cross section of DCX has been measured in the reaction
$A(\pi^-,\pi^+)X$ for $^6Li$ and $^{16}O$ nuclei and at energies
$T_0=0.6$, $0.75$ and $1.1\;GeV$  ($\langle\theta\rangle\approx5^0$)
[1].  The kinematics of the process was chosen in such a way as
to forbid production of an extra pion, $\Delta T=T_0-T\le m_{\pi}$,
where $T$ is the kinetic energy of the produced pion. In this
respect it differs strongly from the kinematics studied in
Ref.[6]. The energy of recoil nucleons has not been measured and it
is necessary to integrate over an energy of the produced neutrons in
the kinematical region indicated above and to average over Fermi
motion of the protons participating in the reaction (Fig.1). Note
that the masses of the intermediate hadronic states, $H$, which can be
produced in the reaction $\pi^-p\to Hn$, increase with incident
energy as $M^2_H\sim2m_NE_0$ (even in the kinematical region of the
experiment).

\begin{figure}[bhtp]
\begin{center}
\vspace*{1.0cm}
\mbox{\epsfig
{figure=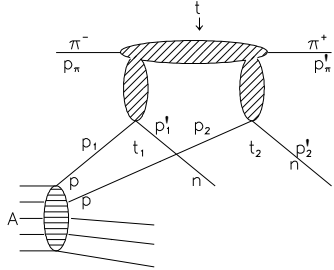,width=11cm,
            bbllx=0,bblly=0,bburx=567,bbury=310}}
\caption [] {Pion double charge exchange on nucleus (the most general
diagram).}
\label{diagram2}
\end{center}
\end{figure}

\begin{figure}[bhtp]
\begin{center}
\vspace*{1.0cm}
\mbox{\epsfig
{figure=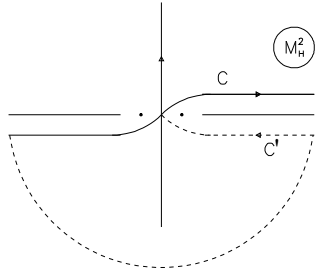,width=11cm,
            bbllx=0,bblly=0,bburx=567,bbury=310}}
\caption [] {Contour of integration in complex $M^2_H$ plane.}
\label{diagram3}
\end{center}
\end{figure}

The most general diagram for pion DCX on nuclei 
is shown in fig.~\ref{diagram2}.
In the Gribov approach [5] an integration over the longitudinal
momenta of nucleons $p_i$, $p'_i$ ($i=1,2$) in 
fig.~\ref{diagram2} can be rewritten
as an integral over $M^2_H=(p_{\pi}+p_1-p'_1)^2=(p'_{\pi}+p'_2-
p_2)^2$, square of invariant mass of the intermediate hadronic
system, which is taken along the contour $C$ in the complex
$M^2_H$ plane, shown in fig.~\ref{diagram3}. 
The integral at large $M^2_H$ is
effectively cut due to a smallness of the Fermi momentum and
experimental limits on kinetic energies of final nucleons
$T'_1+T'_2\le\Delta T$ ($M^2_H\la2E_0\,(\Delta T\cdot m_N)^{1/2}$).
At large enough energies it is convenient to deform the contour $C$ to
$C'$ (fig.~\ref{diagram3}) 
and to drop a contribution of a large circle if the
amplitude decreases with $M^2_H$ faster than $1/M^2_H$ at large
$M^2_H$. The asymptotic behaviour at large $M^2_H$ is determined by
the corresponding $j$--plane singularity in the $t$ channel. In our
case it corresponds to an exchange of a state with the charge $Q=2$
(isospin $I\ge2$). There are no known Regge poles with such "exotic"
quantum numbers and it is reasonable to assume that the amplitude
decreases fast enough at large $M^2_H$. In this case the integration
over $C'$ corresponds to taking a discontinuity over the physical cut
on a real axis. In this way we get the diagrams of 
fig.~\ref{diagram1}(a),(b) as the
pole contributions and the diagram of 
fig.~\ref{diagram1}(c) represents an
intermediate $2\pi$ state. Contributions from intermediate
$3\pi,\;4\pi\;\ldots$ states should be also taken into account though
their contribution can be small. These considerations, strictly
speaking, are valid at high energies $E_0\gg1\;GeV$. For energies
$E_0\la1\;GeV$ a contribution from a circle in $M^2_H$ plane is not
vanishingly small in general. In our estimates in the following we
shall neglect this contribution.

\begin{figure}[bhtp]
\begin{center}
\vspace*{-1.0cm}
\mbox{\epsfig
{figure=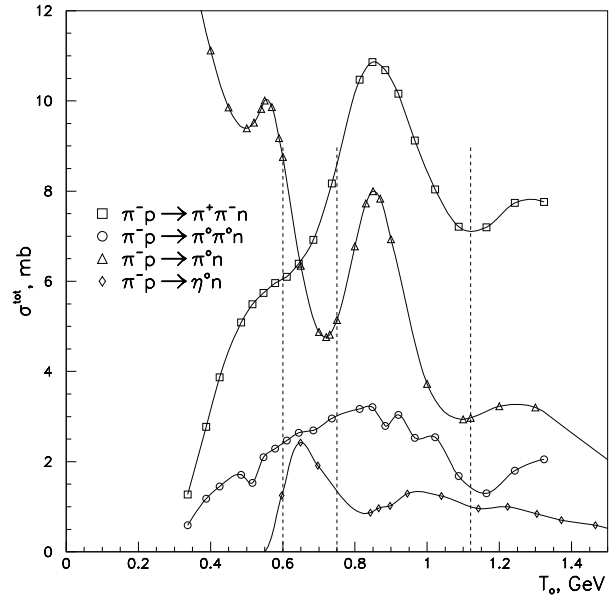,width=9cm,
            bbllx=0,bblly=0,bburx=567,bbury=310}}
\caption [] {Energy dependence for total cross sections of
$\pi^-p\to Hn$ processes. Vertical dashed lines mark three 
values of the incident $\pi^-$ energy in the experiment [1].          }
\label{diagram4}
\end{center}
\end{figure}

Experimental data [7] on cross sections of different pion
charge--exchange processes on the proton in the energy region
$0.3\;GeV<T_0<1.3\;GeV$ are compiled in fig.~\ref{diagram4}. 
It follows from this
figure that $\eta^0$ production gives a relatively small correction
to the diagram of fig.~\ref{diagram1}(a), 
while production of two pions is a very
important competing mechanism (especially for $T_0>0.5\;GeV$).

Note that for forward scattering on nuclei the inelastic corrections
to Glauber model are important only at high energies $E_0\ga5\;GeV$.
They are not very large and are approximately equal $\sim20\div30\%$
of elastic correction [8] (but important for agreement with
experiment).  It is interesting that in the DCX case inelastic
rescatterings can be dominant already at much smaller energies
$T_0\ga1\;GeV$.

\begin{figure}[bhtp]
\begin{center}
\vspace*{3.0cm}
\mbox{\epsfig
{figure=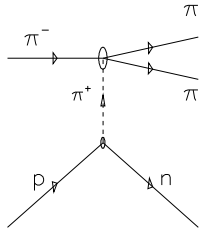,width=11cm,
            bbllx=0,bblly=0,bburx=567,bbury=310}}
\caption [] {The process $\pi^-p\to\pi\pi n$ in one pion exchange
model.}
\label{diagram5}
\end{center}
\end{figure}

Now we shall estimate the contributions of inelastic rescatterings
(see Tables 1,2 below) using experimental data on particle production
in $\pi^-p$ interactions.  Diagrams of 
fig.~\ref{diagram1} in general interfere in
cross section. However if the $\pi\pi$--production amplitude is
dominated by $\pi$ exchange 
(fig.~\ref{diagram5}) then there will be no
interference between diagrams of 
fig.~\ref{diagram1}(a) and fig.~\ref{diagram1}(c) due to difference
in spin structure of nucleon vertices.  Thus in the following we
shall neglect interference terms and shall write the forward cross
section of DCX in the following form

\be
\frac{d\sigma_{DCX}}{d\Omega}=
\frac{d\sigma^{\pi^0}_{DCX}}{d\Omega}+
\frac{d\sigma^{\eta^0}_{DCX}}{d\Omega}+
\frac{d\sigma^{\pi^+\pi^-}_{DCX}}{d\Omega}+
\frac{d\sigma^{\pi^0\pi^0}_{DCX}}{d\Omega}
\ee
where the cross section $d\sigma^H_{DCX}/d\Omega$ denote square of
amplitude of the diagrams of 
fig.~\ref{diagram1} for an intermediate state $H$.

\begin{figure}[bhtp]
\begin{center}
\vspace*{-1.0cm}
\mbox{\epsfig
{figure=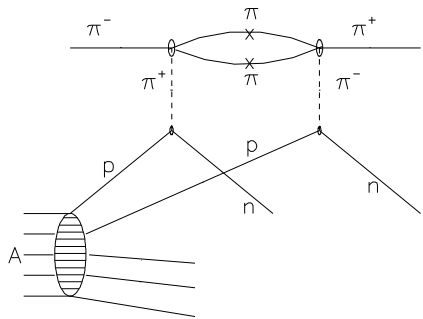,width=14cm,
            bbllx=0,bblly=0,bburx=567,bbury=310}}
\caption [] {Pion double charge exchange in one pion exchange model.}
\label{diagram6}
\end{center}
\end{figure}

\begin{figure}[bhtp]
\begin{center}
\vspace*{4.0cm}
\mbox{\epsfig
{figure=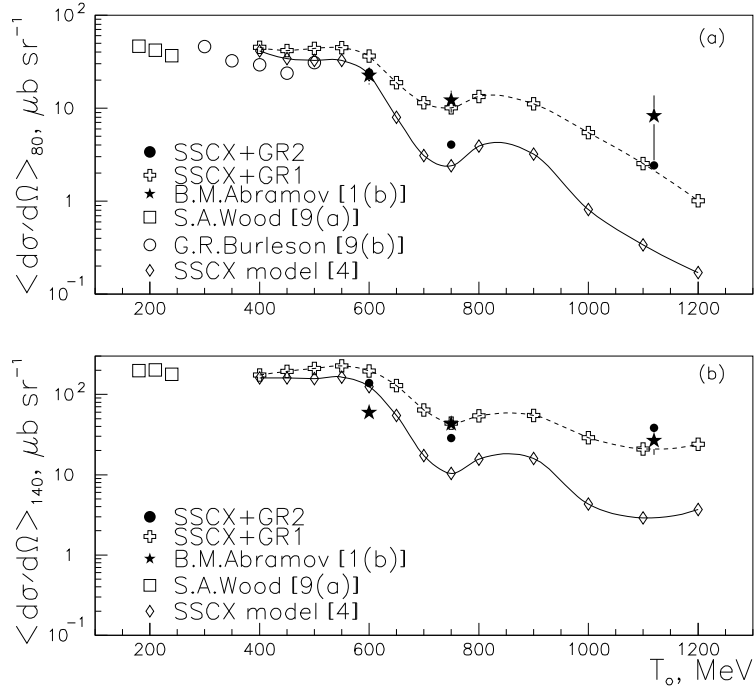,width=11cm,
            bbllx=0,bblly=0,bburx=567,bbury=310}}
\caption [] {Experimental cross sections of the inclusive forward 
pion double charge exchange on $^{16}$O integrated over the region 
$\Delta T$ = 0 -- 80 MeV (a) and $\Delta T$ = 0 -- 140 MeV (b). 
The theoretical calculations are performed in the framework of 
cascade model (SSCX) and with the inelastic Glauber rescattering 
corrections (GR).}
\label{diagram7}
\end{center}
\end{figure}

For single particle intermediate state ($H=\pi^0$, $\eta^0$)

\be
\frac{d\sigma^H_{DCX}}{d\Omega}\sim\left(\int
\frac{d\sigma_{\pi^-p\to Hn}}{dt_1}dt_1\right)^2.
\ee
This is also true for $s$--wave $\pi\pi$--nonresonance intermediate
states, which dominates $\pi\pi$ production at low energies. So we
can write

\be
\frac{d\sigma_{DCX}}{d\Omega}=
\frac{d\sigma^{\pi^0}_{DCX}}{d\Omega}\left(1+\sum_H\Delta_H\right)
\ee
where

\be
\left.
\Delta_H=\left(\int\frac{d\sigma_{\pi^-p\to Hn}}{dt_1}dt_1\right)^2
\right/\left(\int\frac{d\sigma_{\pi^-p\to\pi^0n}}{dt_1}dt_1\right)^2.
\ee
For production of two (or more) pions the quantity, which enters
into (4), in general does not coincide with the cross section of the
corresponding reaction integrated over the kinematical variables of
pions because we consider nondiagonal transition $\pi^-\to\pi^+$.
This is especially clear from the diagram of 
fig.~\ref{diagram1}(c) in the pion
exchange model, i.e. fig.~\ref{diagram6}. 
In this case an imaginary part of
backward $\pi^-\pi^+$ amplitude (not forward elastic $\pi^-\pi^+$
amplitude) enters into the diagram. A difference between forward and
backward amplitudes is connected to an interference between $\pi\pi$
states with odd and even angular momenta and can be taken into
account if a phase shift analysis is used. However this interference
is small for masses of $\pi\pi$ states $\la600\;MeV$, which give a
dominant contribution at the investigated region of energies. The
contribution of $\pi^0$ into $d\sigma_{DCX}/d\Omega$
($d\sigma^{\pi^0}_{DCX}/d\Omega$ in Eqs.(1),(3)) has been
calculated before in Ref.[1] for $^{16}O$ (see Table 2 and 
abbreviation $SSCX$ model in fig.~\ref{diagram7}), 
using a Monte Carlo cascade 
model [4], developed to study pion induced multichannel reactions 
at pion energies above 0.4 GeV.

Corrections $\Delta_H$ due to different intermediate states
calculated using experimental data [7] on corresponding
processes are given in Table 1 and their total contribution is
shown in fig.~\ref{diagram7}. 
It follows from Table 1 that the main contribution
into DCX is given by the process $\pi^-p\to\pi^+\pi^-n$. An account
of all corrections $\sum_H\Delta_H$ ($SSCX+GR1$ in 
fig.~\ref{diagram7}) leads to a
reasonable agreement with the experiment (except values at
$T_0=0.6\;GeV$ where these corrections are less reliable).

\bigskip

\bc
\bt{l|lllllllll}
$T_0,\;GeV$&0.4 &0.5 &0.6 &0.7 &0.8 &0.9 &1.0 &1.1
&1.2\\[0.15cm]\hline
 & & & & & & & & & \\
$\Delta_{\eta}$      &--  &--  &0.02&0.15&0.03&0.02&0.12&0.14&0.09\\[0.25cm]
$\Delta_{\pi^0\pi^0}$&0.01&0.03&0.08&0.32&0.21&0.17&0.47&0.30&0.21\\[0.25cm]
$\Delta_{\pi^+\pi^-}$&0.08&0.32&0.48&2.22&2.21&2.28&5.18&6.00&5.32\\[0.25cm]\hline
 & & & & & & & & & \\
$\sum_H\Delta_H$&0.09&0.35&0.58&2.69&2.45&2.47&5.77&6.44&5.62\\ \hline
\et

\vspace{0.4cm}

\parbox{15cm}{{\bf Table 1}. The quantities $\Delta_H$ (see
Eqs.(3),(4)) of the inelastic rescatterings for $H=\eta^0$,
$\pi^0\pi^0$, $\pi^+\pi^-$.}

\ec
\vspace{0.5cm}

In this simple estimate we have not yet taken into account
experimental limitations on energy transferred to produced neutrons
and to spectator nucleons, 
$\Delta T^{\max}=140\;$ (or $80)\;MeV$, used in the experiment [1].
This condition influences the regions of integrations on
$t_1$ and $t_2$ (see fig.~\ref{diagram2}) 
in Eq.(4) and can lead to some distortion
of the results obtained above if the reaction $\pi^-p\to\pi^0n$ and
other processes $\pi^-p\to Hn$ have different $t$ dependences. We
have estimated an importance of this effect using the following
simple model:

{\bf a)} We considered a configuration when final neutrons have
approximately equal kinetic energies $T'_1\simeq T'_2\le\Delta T^{\max}/2$.
This leads to the following limits on invariant momentum transfer
$|t_i|\le2m_N\left(\Delta T^{\max}/2\right)$.

{\bf b)} The most important intermediate states
$\pi^+\pi^-$ and $\pi^0\pi^0$ were considered.

{\bf c)} The quantities $\Delta_{\pi^+\pi^-}$ and
$\Delta_{\pi^0\pi^0}$ in Eqs.(3),(4) were replaced by

$$
\left.
\Delta'_{\pi^+\pi^-}=\left(\int dM
\int\limits^{t^{\exp}_{1\max}}_{t_{1\min}(M)}
\frac{d^2\sigma_{\pi^+\pi^-}(M,t_1)}{dM\,dt_1}dt_1\right/
\int\limits^{t^{\exp}_{1\max}}_{0}(d\sigma_{\pi^0}/dt_1)dt_1\right)^2,
\eqno{(5a)}
$$
$$
\left.
\Delta'_{\pi^0\pi^0}=\left(\sigma^{tot}_{\pi^0\pi^0}\right/
\sigma^{tot}_{\pi^+\pi^-}\right)^2\cdot\Delta'_{\pi^+\pi^-}
\eqno{(5b)}
$$

The quantity $\Delta'_{\pi^+\pi^-}$ has been calculated using
experimental data on distributions on $\cos\theta(n,p)$ and
$M^2(\pi^+\pi^-)$ for the reaction $\pi^-p\to\pi^+\pi^-n$ given in
Ref.[7c] for the interval of energies we are interested in.
Differential cross section $d\sigma/dt$ for the reaction
$\pi^-p\to\pi^0n$ has been taken from the result of the phase shift
analysis (see [7(d)]). The results are given in Table 2 
and in fig.~\ref{diagram7}
($d\sigma_{DCX}/d\Omega$ and $d\sigma'_{CDE}/d\Omega$ in Table 2 and
$SSCX+GR2$ in fig.~\ref{diagram7}). 
We see that taking into consideration the
experimental limitations on $\Delta T$ leads to substantial decrease
of rescattering correctons at $T_0=0.6\;GeV$ and their increase at
$1.12\;GeV$.

\bigskip

\bc
\bt{l|lll}
$T_0,\;GeV$                    &0.6&0.75&1.12\\[0.15cm]\hline
 & & &\\
$d\sigma^{\pi^0}_{DCX}/d\Omega,\,\mu b/sr$&
125.0 (23.0)&10.4 (2.4)&3.1 (0.31)\\[0.25cm]
$1+\sum_H\Delta_H$             &1.58  &4.21 &7.28\\[0.25cm]
$d\sigma_{DCX}/d\Omega,\,\mu b/sr$&
197.5 (36.3)&43.8 (10.1)&22.6  (2.3)\\[0.25cm]
$1+\Delta'_{\pi^+\pi^-}+\Delta'_{\pi^0\pi^0}$&1.11 (1.05)&2.75
(1.69)&12.3 (7.81)\\[0.25cm]
$d\sigma'_{DCX}/d\Omega,\,\mu b/sr$&139.0
(24.1)&28.6 (4.06)&38.3 (2.42)\\[0.25cm] \hline
 & & &\\
$d\sigma^{\exp}/d\Omega,$
&$59.6\pm7.4$&$43.3\pm5.5$&$26.6\pm8.9$\\[-0.15cm]
$\mu b/sr$&$(22.6\pm4.5)$&$(12.2\pm3.2)$&$(8.3\pm5.5)$\\
\hline
\et

\vspace{0.4cm}

\parbox{15cm}{{\bf Table 2}. The cross section of the inclusive pion
DCX on $^{16}O$ measured in experiment [1b]
($d\sigma^{\exp}/d\Omega$) and calculated:

(a) in Ref.[1] for $H=\pi^0$ using an algorithm of
Ref.[4] ($d\sigma^{\pi^0}_{DCX}/d\Omega$);

(b) according to Eqs.(1)--(4) for $H=\pi^0$, $\eta^0$, $\pi^0\pi^0$,
$\pi^+\pi^-$ ($d\sigma_{DCX}/d\Omega$);

(c) according to Eqs.(1)--(3),(5) for $H=\pi^0$, $\pi^0\pi^0$,
$\pi^+\pi^-$ taking into account the kinematical region $\Delta T$ of
the experiment [1] ($d\sigma'_{DCX}/d\Omega$).

All the differential cross sections presented are integrated over the
two regions of $\Delta T$: from 0 to $140\;MeV$ and to $80\;MeV$ (in
parentheses).}

\ec
\vspace{0.5cm}

An overall agreement between experimental data and predictions of
the model with inelastic rescatterings is quite satisfactory taking
into account that we have made several simplifying assumptions
discussed above. In particular in the region of energies
$T_0\la1\;GeV$ there can be a substantial contribution from the
production of baryon resonances in the direct channel of the process
$\pi^-p\to\pi^+\pi^-n$ and an interference between contributions of
different diagrams of fig.~\ref{diagram1} is possible.

\begin{figure}[bhtp]
\begin{center}
\vspace*{-1.0cm}
\mbox{\epsfig
{figure=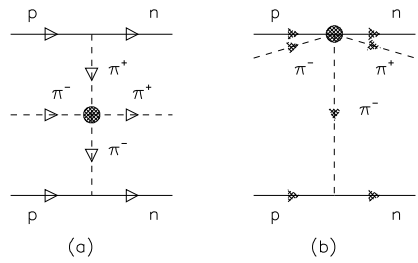,width=16cm,
            bbllx=0,bblly=0,bburx=567,bbury=310}}
\caption [] {Pion double charge exchange in the meson exchange
currents mechanism.}
\label{diagram8}
\end{center}
\end{figure}

Let us discuss a relation between inelastic rescatterings considered
in our paper and other mechanisms which have been proposed for the
DCX process. One of the most popular mechanisms is a model of
meson exchange currents [10], which corresponds to the diagrams
of fig.~\ref{diagram8}
for $\pi^-pp\to\pi^+nn$ transition. The black dots in this
figure are point--like vertices. Comparing the diagram of 
fig.~\ref{diagram8}(a) with
the diagram of fig.~\ref{diagram6}
we see that both take into account
$\pi^-\pi^+\to\pi^+\pi^-$ transitions, but the diagram of 
fig.~\ref{diagram8}
takes into account a real part of this amplitude, which is small for
soft pions, while the diagram of fig.~\ref{diagram6}
is related to an imaginary
part of this amplitude, which is close to a maximum allowed by
unitarity in the region of $M_{\pi\pi}\sim500-800\;MeV$. Inclusion of
the diagram of fig.~\ref{diagram8}(b)
leads to an extra decrease of the total
contribution of this mechanism. A recent calculation for the exclusive
DCX reaction [11], which takes into account, besides the diagram
of fig.~\ref{diagram1}(a),  
the meson exchange currents (fig.~\ref{diagram8}) 
shows that the
contribution of the last mechamism is too small to account for the
experimentally observed cross sections of DCX at energies above
$0.5\;GeV$.

\begin{figure}[bhtp]
\begin{center}
\vspace*{1.0cm}
\mbox{\epsfig
{figure=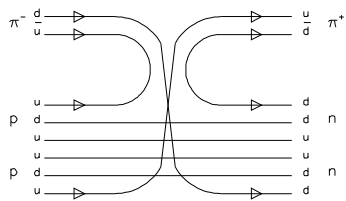,width=14cm,
            bbllx=0,bblly=0,bburx=567,bbury=310}}
\caption [] {Quark diagram for pion double charge exchange.}
\label{diagram9}
\end{center}
\end{figure}

Another possible mechanism for DCX is an interaction of a pion with a
correlated pair of nucleons (see e.g. [12]), which can be
represented at the quark level by the diagram of 
fig.~\ref{diagram9}. The
characteristic feature of such mechanism (as well as of meson
exchange currents), is that only neighboring nucleons of a nucleus can
interact in this way.  On the contrary, inelastic rescatterings of
fig.~\ref{diagram1}(b),(c)
can involve nucleons separated by large longitudinal
distances (at least at high energies).  This leads to a difference in
$A$ dependence of the DCX processes for mechanisms mentioned above.

If our assumption that integration over $M^2_H$
converges fast enough is correct then at energies much higher than
$1\;GeV$ the cross section of DCX will be governed by the
corresponding $\rho$ and $\pi$ exchanges in $t_i$ channels. In this
case the cross section should decrease at least as fast as
$s^{2[(2\alpha_{\rho}(0)-1)-1]}\approx s^{-2}$ (see, e.g. review
lectures [13]) and its rather slow energy dependence at energies
below $1.1\;GeV$ should change to a much faster decrease.  What can
be expected in the region of energies $T_0\sim2-5\;GeV$, the closest
one to the experiment [1]?  The theoretical estimates used above
become more reliable at these energies because corrections to
theoretical formulae decrease with energy and the one pion exchange
model gives an adequate description of the main features of
$\pi\to2\pi$ process here [14]. The available data on the total cross
section of the reaction $\pi^-p\to\pi^+\pi^-n$ [15] and
$\pi^-p\to\pi^0n$ [7(d)] show that the value of inelastic
rescatterings at these energies is still rather large
($\Delta_{\pi^+\pi^-}\gg1$). Thus we expect large deviations from the
standard DCX mechanism also in this energy region. The effect can be
estimated using the method proposed in this paper.

It is highly
desirable to extend measurements of DCX to these energies taking into
account that DCX gives a unique possibility to observe a large
contribution of inelastic rescatterings and to understand
better a mechanism of this process. (This is in contrast to the  case
of elastic scattering of hadrons on nuclei where inelastic
rescatterings lead to rather small effects.) The value of the
inclusive pion DCX cross section integrated over the region $\Delta
T=0\div140\;MeV$ is expected to be high enough ($\sim20\;\mu b/sr$
at $2\;GeV$) to be detected in experiment. The experimental study of
the $A$ dependence could give additional information in order to
distinguish between the sequential mechanism and the pion interaction
with correlated pairs of nucleons.

\section*{Acknowledgments}

Authors are thankful to K.G.Boreskov for discussions and to
V.V.Kulikov for reading the manuscript and remarks. A.P.K. is also
indebted to I.S. and I.I.Tsukerman for current support. The work of
A.B.K. was supported in part by RFBR Grant No.96--02--19184a.


\clearpage

\end{document}